\documentclass{webofc}

\usepackage[varg]{txfonts}   
\usepackage{hyperref}
\usepackage{url}
\hypersetup{colorlinks=true,citecolor=blue,urlcolor=blue,linkcolor=blue}
\begin{document}
\title{Origin of the electric hexadecapole isomer in $^{93}$Mo}
%
%

\author{\firstname{B.} \lastname{Maheshwari}\inst{1}\fnsep\thanks{\email{bhoomika.physics@gmail.com}} \and
        \firstname{P.} \lastname{Van Isacker}\inst{1} \and
        \firstname{P. M.} \lastname{Walker}\inst{2}}

\institute{Grand Accélérateur National d'Ions Lourds,
CEA/DSM-CNRS/IN2P3, Bvd Henri Becquerel, BP 55027, F-14076 Caen, France 
\and
Department of Physics, University of Surrey, Guildford, GU2 7XH, United Kingdom 
          }

\abstract{We present a shell-model analysis of $^{93}$Mo to investigate the unusual behavior of its ${21/2}^+$ isomer -- a prominent candidate for nuclear excitation by electronic capture. This state is unique as its decay is dominated by a slow electric hexadecapole $E4$ transition, while the typically much faster electric quadrupole $E2$ decay path is energetically forbidden. We investigate the microscopic origin of this phenomenon by examining in detail the structure of the wave functions of the initial and final states, and the $E4$ transition matrix elements. This analysis of $^{93}$Mo is contrasted with that of its particle-hole conjugate, $^{99}$Cd, where such an $E4$ transition is absent. 
}
\maketitle

\section{Introduction}
Nuclear states can decay from a higher to a lower energy level by emitting photons, a process governed by the multipolarity of the transition. While magnetic dipole $(M1)$ and electric quadrupole $(E2)$ transitions are most common, transitions of higher multipolarity are significantly suppressed. This suppression can lead to the formation of long-lived excited states, known as nuclear isomers~\cite{isomer}. Such high-multipolarity transitions are extremely sensitive to the radial and angular overlaps of the wave functions of the initial and final states, making them excellent probes of nuclear structure. 

This work focuses on the remarkable $21/2^+$ isomer in $^{93}$Mo which decays via an electric hexadecapole $(E4)$ transition. This isomer has garnered significant attention as the only isomer for which nuclear excitation by electron capture (NEEC) has been reported~\cite{chiara2018}, a claim that remains under active investigation~\cite{guo2022}. The NEEC process relies on the precise energies and transition probabilities of the nuclear states. In $^{93}$Mo, the idea involves a 5 keV trigger to induce a transition from the long-lived $21/2^+$ isomer (half-life 6.85 hours) to the short-lived $17/2^+$ isomer (half-life 3.53 nanoseconds), potentially releasing stored energy with an energy gain of at least 50 times~\cite{carroll2024}. Understanding the nuclear structure that enables this scheme is therefore paramount.   

\begin{figure*}[!htb]
\centering
\resizebox{0.65\textwidth}{!}{\includegraphics[width=\textwidth]{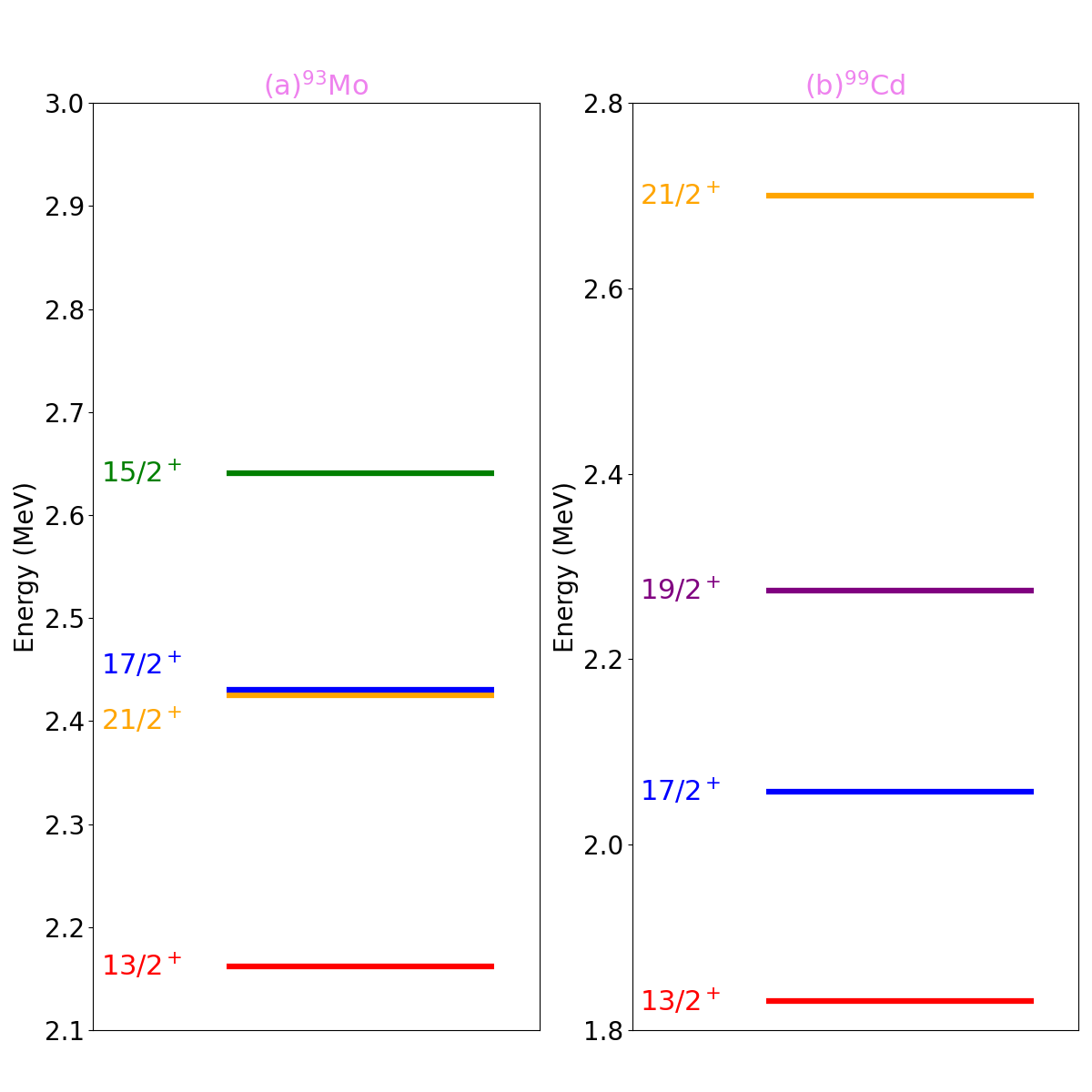}}
\caption{(Color online) A part of the experimental level scheme in (a) $^{93}$Mo and (b) $^{99}$Cd. The ${21/2}^+$ state in $^{93}$Mo can only decay via possible $E4$ transition to the lower-lying ${13/2}^+$ state.}
\label{fig:en}\end{figure*}

The existence of the $E4$ isomer in $^{93}$Mo is due to an unusual inversion of the $21/2^+$ and $17/2^+$ states as shown in the figure~\ref{fig:en}. In the simple seniority scheme for three protons in the $0g_{9/2}$ orbital, as seen in $^{93}$Tc, these states are ordered in energy by spin. However, in $^{93}$Mo, the configuration is predominantly $\nu 1d_{5/2} \otimes \pi 0g_{9/2}^2  $, and  the strong attractive interaction between the neutron and the proton pair inverts the level ordering, placing the $21/2^+$ state below the $17/2^+$ state~\cite{auerbach1964}. This inversion energetically forbids the fast $21/2^+ \rightarrow 17/2^+$ $E2$ transition, forcing the $21/2^+$ state to decay via the slow $E4$ transition to the lower-lying $13/2^+$ state. 

While large-scale shell-model calculations have reproduced this feature~\cite{hasegawa2011}, a transparent, quantitative analysis of the underlying mechanism is often obscured. We recently showed the quantitative role of neutron-proton $(\nu\pi)$ interaction responsible for this inversion and provided a revised smaller estimate for the crucial $B(E2; 17/2^+ \rightarrow 21/2^+)$ value relevant to NEEC~\cite{maheshwari2025}. Here, we extend that work to provide an explicit microscopic origin of the $E4$ transition itself. We analyze the specific wave function components and couplings that give rise to the $E4$ transition probability in $^{93}$Mo. This is found to be in contrast to its particle-hole conjugate, $^{99}$Cd, where such an inversion of the $17/2^+$ and $21/2^+$ states does not occur (see figure~\ref{fig:en}) and consequently, no $E4$ isomer is observed. 

\section{Formalism}
\label{sec:formalism}
The shell-model Hamiltonian is written as
\begin{equation}
\hat H=\sum_i\epsilon_i\hat n_i+\frac{1}{4}\sum_{i j k l}V_{ijkl} a^\dagger_i a^\dagger_j a_l a_k +\cdots,
\label{eq:ham}
\end{equation}
where the first term contains the single-particle energies and the second term represents the two-body residual interaction. We perform calculations for $^{93}$Mo within the $\nu1d_{5/2}\otimes\pi0g_{9/2}^2$ model space. The two-body matrix elements for the like-particle interaction $ V_{\pi\pi}$ are derived empirically from the spectrum of $^{92}$Mo, which has two valence protons in the $0g_{9/2}$ orbital. The adopted values are $V_{\pi\pi}(J=0) =0.000$, $ V_{\pi\pi}(J=2) =1.510$, $ V_{\pi\pi}(J=4) =2.283$, $ V_{\pi\pi}(J=6) =2.612$, and $ V_{\pi\pi}(J=8) = 2.761$ (all in MeV). Similarly, the $\nu\pi$ two-body interaction matrix elements are fixed from the energy spectrum of $^{92}$Nb, corresponding to the coupling of a neutron $\nu1d_{5/2}$ with a proton $\pi0g_{9/2}$. The values are $  V_{\nu\pi}(J=2) = 0.135$, $ V_{\nu\pi}(J=3) =0.286 $, $ V_{\nu\pi}(J=4) =0.480$, $ V_{\nu\pi}(J=5) =0.357$, $ V_{\nu\pi}(J=6) =0.501$, and $ V_{\nu\pi}(J=7) =0.000$ (all in MeV). The interaction ${V}_{\nu\pi}^J$ is the strongest for the fully aligned state with $J=7$,
which is the ground state of $^{92}$Nb. For each set of matrix elements, we have set the most attractive one to zero. This is justified when the focus is solely on excitation energies rather than absolute energies. The same applies to the single-particle energies for our present configuration, which includes only one neutron and one proton orbital; they have no impact on the computed excitation energies. The calculations are performed with the Mathematica code {\tt shell.m}~\cite{isackercode} within the $J$-scheme using coefficients of fractional parentage~\cite{talmi}. The eigenvalue problem, however, requires a numerical diagonalization.

The reduced probability for an electric transition of multipolarity $L$, between an initial state $\alpha_{\rm i}J_{\rm i}$ and a final state $\alpha_{\rm f}J_{\rm f}$, is given by 
\begin{equation}
B(EL;\alpha_{\rm i}J_{\rm i}\rightarrow\alpha_{\rm f}J_{\rm f})=
\frac{|\langle\alpha_{\rm f}J_{\rm f}||\hat O(EL)||\alpha_{\rm i}J_{\rm i}\rangle|^2}{2J_{\rm i}+1},
\label{eq:bela}
\end{equation}
The electric transition operator $\hat O(ELM)=\sum_ke_kr^L_kY_L^M(\theta_k,\phi_k)$ depends on the effective charge, and contains a radial part $r^L$ and an angular part described by the spherical harmonic $Y_L^M(\theta,\phi)$. The reduced matrix element $\langle\cdot||\cdot||\cdot\rangle$~\cite{talmi} is independent of the projection quantum numbers of the angular momentum.
Equation~(\ref{eq:bela}) applies to many-body eigenstates $\alpha J$,
obtained after diagonalization of the Hamiltonian matrix,
which are expanded in a chosen shell-model basis such that 
\begin{align}
|\alpha_{\rm i}J_{\rm i}M_{\rm i}\rangle={}&
\sum_k a_k(\alpha_{\rm i}J_{\rm i})|kJ_{\rm i}M_{\rm i}\rangle,
\nonumber\\
|\alpha_{\rm f}J_{\rm f}M_{\rm f}\rangle={}&
\sum_l b_l(\alpha_{\rm f}J_{\rm f})|lJ_{\rm f}M_{\rm f}\rangle,
\label{eq:expan}
\end{align}
where $k$ ($l$) labels basis states with angular momentum $J_{\rm i}$ ($J_{\rm f}$).
Combination of Eqs.~(\ref{eq:bela}) and~(\ref{eq:expan})
leads to a generalized expression for the reduced transition probability,
\begin{align}
&B(EL;\alpha_{\rm i}J_{\rm i}\rightarrow\alpha_{\rm f}J_{\rm f})=
\frac{1}{2J_{\rm i}+1}
\biggl|\sum_{kl}a_k(\alpha_{\rm i}J_{\rm i})b_l(\alpha_{\rm f}J_{\rm f})\langle lJ_{\rm f}
||\hat O(EL)||kJ_{\rm i}\rangle\biggr|^2.
\label{eq:belb}
\end{align}
This can explain the interference effects between neutrons and protons.

\begin{figure*}[!htb]
\centering
\resizebox{0.8\textwidth}{!}{\includegraphics[width=\textwidth]{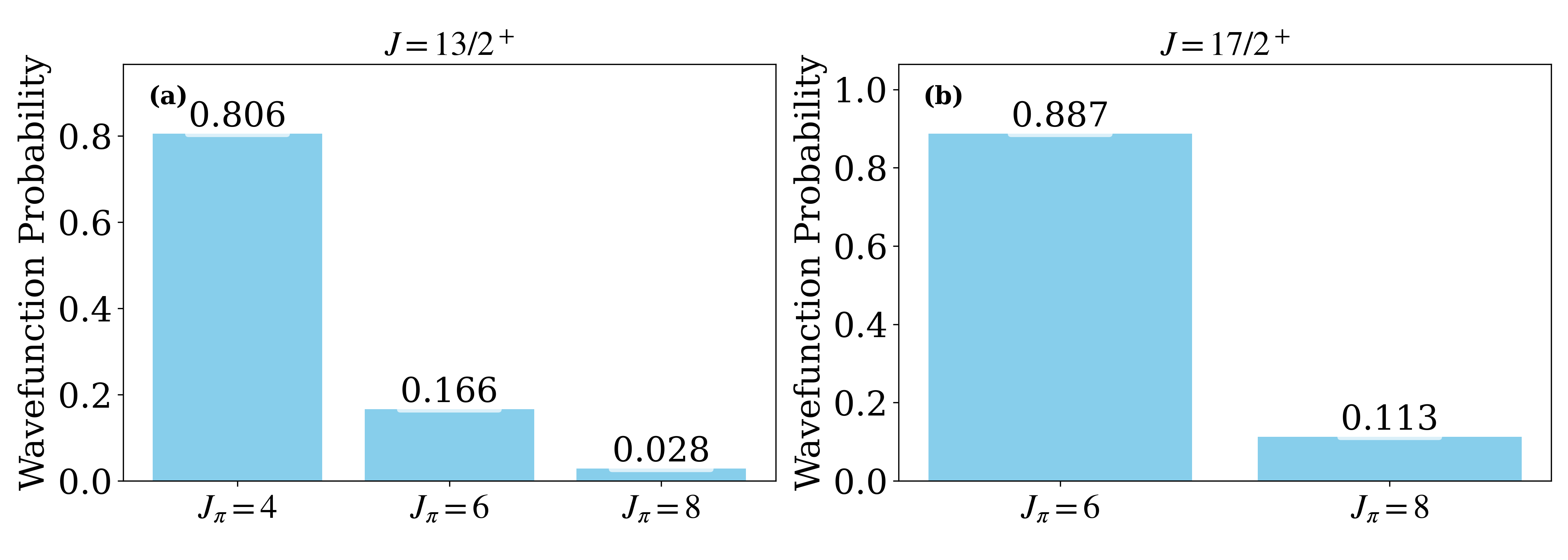}}
\caption{(Color online) Calculated wave-function probability for the lowest-lying (a) ${13/2}^+$ and (b) ${17/2}^+$ states. $J_\pi$ denotes the angular momentum of two protons in the $0g_{9/2}$ orbital which couples to the odd-neutron in the $1d_{5/2}$ orbital to generate the respective states. }
\label{fig:wf}\end{figure*}

In $^{93}$Mo,
there is only one neutron in the $1d_{5/2}$ orbital which constitutes the reduced matrix element of the neutron part of the $EL$ operator. The reduced matrix element of the $EL$ operator between two-proton states in the $0g_{9/2}$ orbital involves the single-particle proton reduced matrix element. Note that this matrix element not only depends on the single-particle angular momentum $j$ but also on the principal quantum number $\cal N$ and the orbital angular momentum $l$. The single-particle reduced matrix elements for the relevant orbitals are calculated using standard expressions~\cite{talmi}. For the $E4$ transition of interest, these are:
\begin{align}
\langle\nu1d_{5/2}||\hat O_\nu(E4)||\nu1d_{5/2}\rangle={}&\frac{513}{4}\sqrt{\frac{1}{7\pi}}e_\nu b^4,
\nonumber\\
\langle\pi0g_{9/2}||\hat O_\pi(E4)||\pi0g_{9/2}\rangle={}&\frac{9}{4}\sqrt{\frac{715}{\pi}}e_\pi b^4,
\label{eq:e4sp}
\end{align}
where $e_\nu$ and $e_\pi$ are the neutron and proton effective charges, and $b$ is the harmonic oscillator length parameter,  $b^2=41.46/(45A^{-1/3}-25A^{-2/3})$~${\rm fm}^2$~\cite{blomqvist1968},
resulting in $B(E4)$ values in units of $e^2{\rm fm}^{8}$.
For $E2$ transitions, the proton effective charge $e_\pi$ is set to 1.32~\cite{ley2023},
and three neutron effective charges $e_\nu$ are adopted:
a) 1.18 from the quadrupole moment $Q(5/2^+)$ of the ground state of $^{91}$Zr~\cite{nndc},
b) 1.73 from the $B(E2;2_1^+\rightarrow0_1^+)$ value in $^{92}$Zr~\cite{nndc},
and c) the averaged value 1.48. For the $E4$ transition, we use the proton effective charge $e_\pi=1.20$, determined from the measured $B(E4; 4^+ \rightarrow 0^+)$ value in $^{92}$Mo~\cite{nndc, thesis} while since no further details are available regarding the neutron, $e_\nu$ is kept the same as for the $E2$ operator. The reduced transition probabilities, $B(E2)$ and $B(E4)$ values, are often expressed in single-particle units, also called Weisskopf units (W.u.), $
B_{\rm W}(E2)=0.0594\times A^{4/3}e^2{\rm fm}^4,$ and $
B_{\rm W}(E4)=0.0628\times A^{8/3}e^2{\rm fm}^8,$
where $A$ is the mass number. 

\section{Discussion}
\label{sec:discussion}

Within the chosen valence space, the coupling of a $1d_{5/2}$ neutron to a $0g_{9/2}^2$ proton pair gives rise to one $21/2^+$, two $17/2^+$, and three $13/2^+$ states. As previously shown~\cite{maheshwari2025} and summarized in Table~\ref{tab:enmo93}, our calculations accurately reproduce the experimental level energies, indicating the crucial inversion with the $21/2^+$ state below the first $17/2^+$ state. This inversion is a direct consequence of the attractive $\nu\pi$ interaction, which is the strongest when spins are maximally aligned. The $21/2^+$ state has a pure structure $|\nu 1d_{5/2} \otimes \pi 0g_{9/2}^2; 21/2 \rangle $, maximizing the alignment. This configuration is strongly lowered in energy pulling it below the $17/2^+$ state. Our calculations show that the lowest $13/2^+$ state is predominantly composed of $J_\pi=4$ proton coupling $(\sim 80\%)$, but with significant admixtures from $J_\pi=6$ and $J_\pi=8$ components, shown in Fig.~\ref{fig:wf}(a). The dominant component for the $17/2^+$ state arises from the less-aligned $J_\pi=6$ proton coupling, shown in Fig.~\ref{fig:wf}(b).  

This situation should be contrasted with the particle-hole conjugate nucleus $^{99}$Cd. Its structure relative to the $^{100}$Sn core can be viewed as one neutron particle in the $1d_{5/2}$ orbital and two proton holes in the $0g_{9/2}$ orbital. According to the Pandya transformation~\cite{pandya1956}, the particle-particle interaction matrix elements are related to the particle-hole ones. Crucially, the attractive neutron-proton interaction that lowers the maximally-aligned state in $^{93}$Mo becomes repulsive for the corresponding particle-hole case. Consequently, no level inversion is expected in $^{99}$Cd. Indeed experimentally the lowest $21/2^+$ state in $^{99}$Cd lies well above the lowest $17/2^+$ state~\cite{nndc}. This could also be reproduced in our calculations~\cite{maheshwari2025}. Hence, the $21/2^+$ can decay by a relatively fast $E2$ transition and no $E4$ isomerism is observed. This comparison starkly illustrates the decisive role of the $\nu\pi$ interaction in creating the conditions for $E4$ isomerism in $^{93}$Mo.

The direct consequence of this level inversion in $^{93}$Mo is that the fast, seniority-allowed $E2$ transition from the $21/2^+$ state to the $17/2^+$ state is energetically forbidden, giving rise to the long-lived isomerism. The reverse transition, $17/2^+ \rightarrow 21/2^+$, is of great importance for the NEEC process. Its strength is dictated by the matrix element between the mixed $17/2^+$ state and pure $21/2^+$ state. The transition is dominated by the quadrupole coupling of the two protons as they change their coupling from $J_\pi=6$ to $J_\pi=8$. The $1d_{5/2}$ neutron contributes a smaller amount via its spectroscopic quadrupole moment. Both contributions interfere constructively, leading to a transition probability proportional to $(0.23\,e_\nu+0.93\,e_\pi)^2$. Our calculated value for this transition is about $40\%$ reduced to a previous estimate~\cite{hasegawa2011}. 

\begin{figure}[!htb]
\centering
\includegraphics[width=0.65\textwidth]{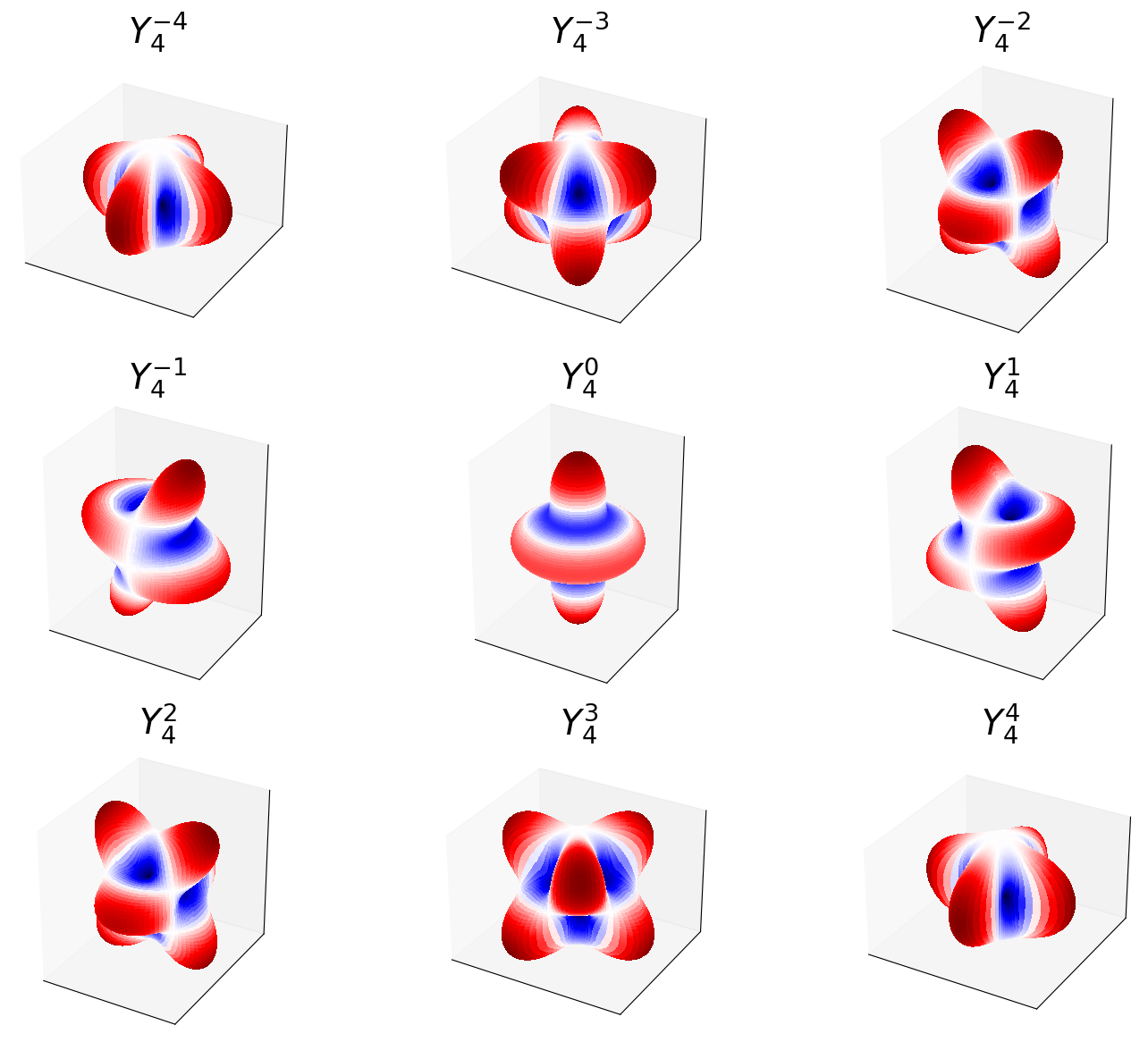}
\caption{(Color online)
\label{fig:yl}
Schematic representation of the real part of the rank-4 spherical harmonics which govern the angular properties of an electric hexadecapole $E4$ transition. The color indicates the sign of the function; red for positive and blue for negative.} 
\end{figure}

\begin{table}[!htb]
\caption{\label{tab:enmo93}
Experimental~\cite{nndc} and calculated energies (in MeV) in $^{93}$Mo. Present results are obtained in the $\nu1d_{5/2}\otimes\pi0g_{9/2}^2$ model space, and also compared with results of Hasegawa {\it et al.}~\cite{hasegawa2011}.}
\centering
\resizebox{0.35\textwidth}{!}{\begin{tabular}{|c|c|c|c|}
\hline
$J^\pi$ & Exp. & Present & Ref.~\cite{hasegawa2011} \\
\hline
${13/2}^+$ & 2.161 & 2.221 & 2.197 \\
    & 2.667 & 2.893 & 2.359 \\
    & --- & 3.036 &--- \\
${17/2}^+$ & 2.429 & 2.454 & 2.398  \\
    & --- & 3.197 & ---  \\
${21/2}^+$ & 2.424 & 2.437 & 2.315 \\
\hline
\end{tabular}}
\end{table}

Furthermore, the observed $E2$ transition from the $17/2^+$ to the $13/2^+$ state serves as a critical benchmark for our calculated wave functions. This transition connects two states with significant configuration mixing. The total transition amplitude is a coherent sum of several components, primarily driven by the proton couplings, $J_\pi=6 \rightarrow J_\pi=4$ and $J_\pi=6 \rightarrow J_\pi=6$. Again, the neutron and proton parts interfere constructively, with the total reduced transition probability proportional to $(0.28\,e_\nu+1.28\,e_\pi)^2$. Our calculated value for this transition, as listed in Table~\ref{tab:emmo93}, is in excellent agreement with the measured value. This provides strong confidence in the accuracy of the wave functions for the chosen nuclear states and related predictions.  

\begin{figure}[!htb]
\centering
\includegraphics[width=0.45\textwidth]{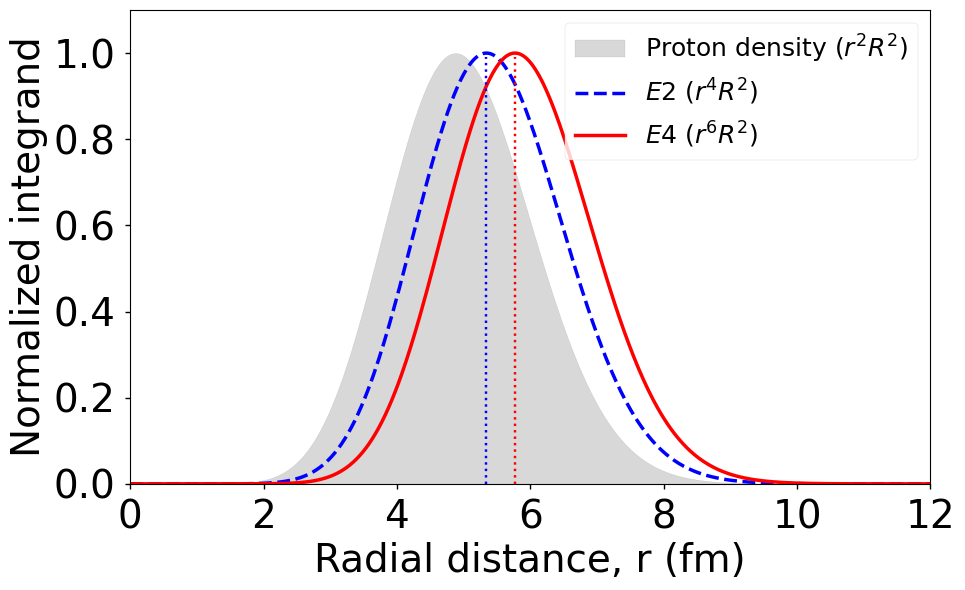}
\caption{(Color online)
\label{fig:rad}
Variation of normalized radial integrands with radial distance involved in $E2$ and $E4$ radial overlaps compared with the proton density shown for the $0g_{9/2}$ orbital in $^{93}$Mo. $R$ denotes the radial harmonic oscillator wave function.} 
\end{figure}

In $^{93}$Mo, with the $E2$ decay path prohibited, the $21/2^+$ isomer must decay to the $13/2^+$ state, requiring an angular momentum change of $\Delta J=4$, mandating an $E4$ transition. The angular momentum character of this transition is governed by the rank-4 spherical harmonics, $Y_4^M (\theta, \phi)$, which are visualized in Fig.~\ref{fig:yl} for the real parts of the characteristic nodal planes and angular symmetries of $Y_4^M (\theta, \phi)$ with $M=-4, -3,... +4$. These functions describe the geometric shape of the $E4$ operator. Each subplot shows a 3D surface where the distance from the center is scaled by the value of the harmonic function. The values are normalized so that the largest lobes are all shown with the same relative size making it easy to compare different $M$ states. Each lobe in the figure represents a region of high probability for the interaction, with the colors (red/blue) indicating the phase (positive/negative), highlighting the nodal patterns and symmetries. For the $21/2^+ \rightarrow 13/2^+$ transition to occur, the participating nucleons must undergo a significant rearrangement of their orbital motion, changing their angular distribution from one that matches the initial state to one that matches the final state. This rearrangement must have an angular pattern that fits one of the complex shapes shown in Fig.~\ref{fig:yl}. In addition, the radial part of the $E4$ operator, with its $r^4$ dependence, makes the transition highly sensitive to the behavior of the wave function at the nuclear surface. Figure~\ref{fig:rad} shows that the radial contributions for $E2$ and $E4$ transitions peak at progressively larger distances, shifting away from the proton's most probable location in the nuclear interior. The strength depends on the radial integrals' $r^L$ dependence quantifying the overlap between the initial and final wave functions, which is weighted heavily towards the nuclear exterior for the $E4$ transition as shown in Fig.~\ref{fig:rad}. Using harmonic wave functions, the relevant single-particle radial integrals for $\langle \nu 1d_{5/2}|r^4| \nu 1d_{5/2} \rangle = 42.75 b^4$, and $\langle \pi 0g_{9/2}|r^4| \pi 0g_{9/2} \rangle = 35.75 b^4$ with $b$ the oscillator length parameter. While both orbitals belong to the same harmonic oscillator shell, the $0g_{9/2}$ orbital is more spatially compact than the $1d_{5/2}$ orbital. The observed $E4$ transition is only possible because both the complex angular rearrangement and the required radial overlap are simultaneously satisfied by the specific admixtures in the shell-model wave functions of the nuclear states. 

\begin{figure}[!htb]
\centering
\resizebox{0.65\textwidth}{!}{\includegraphics[width=\textwidth]{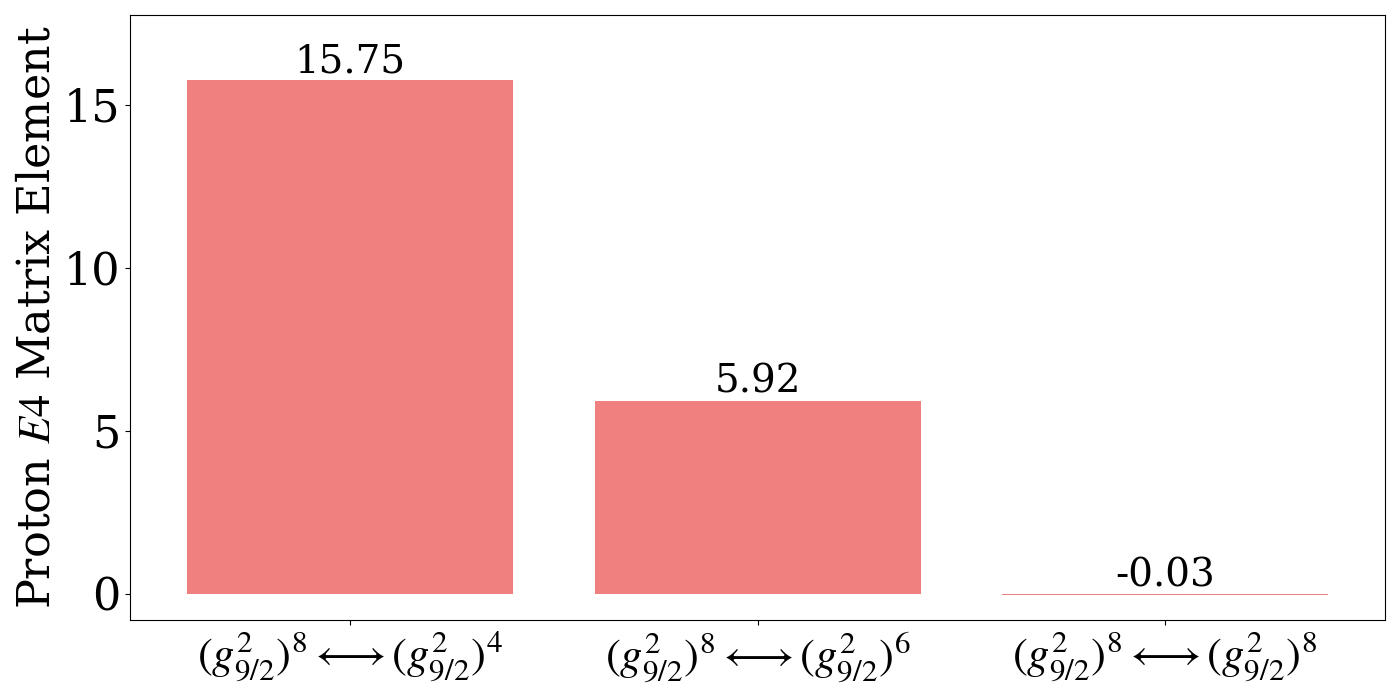}}
\caption{(Color online) Decomposition of the proton contribution to the $E4$ transition matrix element (in units of $b^4$) for the ${21/2}^+ \rightarrow {13/2}^+$ transition in $^{93}$Mo. The bars show the contribution from each two-proton coupling $(J_\pi=4,6,8)$ in the final $13/2^+$ state. The neutron part of the matrix element from the $1d_{5/2} \longleftrightarrow 1d_{5/2}$ transition is calculated to be 23.32 (in units of $b^4$). The constructive interference between proton and neutron contributions leads to a total reduced transition probability of $B(E4;{21/2}^+ \rightarrow {13/2}^+)= (0.83 e_\nu+3.53 e_\pi)^2b^8$.}
\label{fig:e4}\end{figure}

\begin{table}[!htb]
\caption{\label{tab:emmo93}
Experimental~\cite{nndc} and calculated reduced $E2$ and $E4$ transition probabilities in $^{93}$Mo in Weisskopf units (W.u.).
Also listed are the results of Hasegawa \textit{et al.}~\cite{hasegawa2011}.}
\centering
\resizebox{0.7\textwidth}{!}{
\begin{tabular}{|c|c|c|c|c|c|}
\hline
&&
\multicolumn{3}{c|}{Present calculation}&Ref.~\cite{hasegawa2011}\\
$J_{\rm i}\rightarrow J_{\rm f}$&Exp.&
\multicolumn{3}{c|}{$e_\pi=1.32 (E2), 1.20 (E4)$  }&$e_\pi=1.50$\\
\cline{3-5}
&&$e_\nu=1.18$&$e_\nu=1.78$&$e_\nu=1.48$&
$e_\nu=0.50$\\
\hline
${17/2}^+ \rightarrow {21/2}^+$ & --- & 
${2.02}$& $2.40$ & ${2.21}$ & ${3.5}$  \\
${17/2}^+ \rightarrow {13/2}^+$ & {4.48(23)} & {3.67} &{ 4.30} & {3.98} & {4.0} \\
${21/2}^+ \rightarrow {13/2}^+$ & 1.449(17) &  1.25 & 1.50 & 1.37 & 1.9    \\
\hline
\end{tabular}}
\end{table}

The $E4$ transition from the pure $21/2^+$ $(J_\pi=8)$ state can therefore proceed to all three components of the $13/2^+$ state. Fig.~\ref{fig:e4} shows a decomposition of the proton part of the $E4$ matrix element. The largest contribution comes from the $\pi 0g_{9/2}^2 (J_\pi=8) \rightarrow \pi 0g_{9/2}^2 (J_\pi=4)$ component. The phase for $\pi 0g_{9/2}^2 (J_\pi=8) \rightarrow \pi 0g_{9/2}^2 (J_\pi=8)$ stays opposite to the other two proton $E4$ matrix elements. The neutron also contributes via the $1d_{5/2} \rightarrow 1d_{5/2}$ transition. The final expression of the reduced transition probability is found to be $B(E4; 21/2^+ \rightarrow 13/2^+) = (0.83 e_\nu + 3.53 e_\pi)^2 b^8$. The signs of the neutron and proton terms are the same, leading to constructive interference that enhances the transition probability. Table~\ref{tab:emmo93} shows our calculated $B(E4)$ value, which is in excellent agreement with the experimental measurement with a varying neutron effective charge ranging from 1.18 to 1.78. Since neutron $E4$ transition matrix elements are relatively small, this does not influence the final $B(E4)$ value significantly. This confirms that the chosen configuration captures the essential physics of this complex transition.  

\section{Conclusions}

We have performed a detailed quantitative shell-model analysis to elucidate the origin of the $21/2^+$, $E4$ isomer in $^{93}$Mo in the $\nu 1d_{5/2} \otimes \pi 0g_{9/2}^2$ configuration space. The higher-multipole isomerism is a consequence of the level inversion of the $21/2^+$ and $17/2^+$ states due to the strong, attractive $\nu\pi$ interaction in the maximally aligned configuration, $J=7$. A comparison with its particle-hole conjugate nucleus, $^{99}$Cd, where this interaction becomes repulsive, shows no such inversion and no $E4$ isomerism, providing compelling evidence of the proposed mechanism. 

The results for both level energies and transition probabilities show generally good agreement with experimental data especially involving the states of interest, i.e. the $13/2^+$, $17/2^+$ and $21/2^+$ states. Our results indicate a $\sim40\%$ reduction in the theoretical $B(E2)$ value for the $17/2^+ \rightarrow 21/2^+$ transition in $^{93}$Mo compared to the previous estimate~\cite{hasegawa2011}. The $E4; 21/2^+ \rightarrow 13/2^+$ transition probability is also well explained by our calculations. It arises from a coherent, constructive sum of proton and neutron components. We explicitly analyze the $E4$ transition matrix elements, mainly dominated by proton two-body matrix elements in $0g_{9/2}$. This work provides a quantifiable exploration of proton-proton and neutron-proton interactions in order to understand their impact on the formation of peculiar higher-multipole isomers. Such isomers and their detailed understanding are of central importance to ongoing research into induced isomer depletion and NEEC, and to the identification of the similar candidates in other regions of the nuclear chart. For instance, the $M4$ isomer in $^{84}$Rb requires future attention and is presently under study. 
 
\section*{Acknowledgments}
The author BM gratefully acknowledges the financial support from the HORIZON-MSCA-2023-PF-01 project, ISOON, under grant number 101150471. PW acknowledges support from the
UK Science and Technology Facilities Council under Grant
No. ST/V001108/1.

\end{document}